\documentclass[conference]{IEEEtran}

\ifCLASSINFOpdf
\else
\fi
\usepackage{graphicx}
\usepackage{setspace}
\usepackage{amsmath,empheq}
\usepackage{amssymb}
\usepackage{amsthm}
\usepackage{caption}
\usepackage{subcaption}
\usepackage{bm} 
\usepackage{cite}
\usepackage{float}
\usepackage[usenames,dvipsnames]{pstricks}
\usepackage{epsfig}
\usepackage{pst-grad} 
\usepackage{pst-plot} 
\usepackage{tabularx}
\usepackage{flushend}   
\usepackage{amsmath}
\usepackage[T1]{fontenc}

\newcommand{\nn} {\nonumber}
\usepackage{xcolor}
\usepackage{lipsum}
\usepackage{comment}
\usepackage[font=small]{caption}
\allowdisplaybreaks

\allowdisplaybreaks
\usepackage[utf8]{inputenc}

 \IEEEaftertitletext{\vspace{-2.5\baselineskip}}
 \usepackage{svg}

\usepackage{algpseudocode}
\usepackage{algorithm}
\usepackage{xcolor}
\algnewcommand\algorithmicforeach{\textbf{for each}}
\algdef{S}[FOR]{ForEach}[1]{\algorithmicforeach\ #1\ \algorithmicdo} 

\begin{document}
\bstctlcite{IEEEexample:BSTcontrol}

\title{Reinforcement Learning for Secrecy Optimization in Underwater Energy Harvesting Relay Network}



  \author{
    \IEEEauthorblockN{Shalini Tripathi\IEEEauthorrefmark{1}, 
    Ankur Bansal\IEEEauthorrefmark{2},
    and Chinmoy~Kundu\IEEEauthorrefmark{3}} 
      \IEEEauthorblockA{\IEEEauthorrefmark{1}\IEEEauthorrefmark{2}Department of EE, Indian Institute of Technology Jammu, India}
       \IEEEauthorblockA{\IEEEauthorrefmark{3}Wireless Communications Laboratory, Tyndall National Institute, Dublin, Ireland}

}

\maketitle
 \thispagestyle{empty}
\pagestyle{plain}

\begin{abstract}
This paper explores secure communication in an underwater energy-harvesting (EH) relay network that supports hybrid optical-acoustic transmission. The optical hop is modeled using a Gamma–Gamma turbulence channel with pointing errors and may occasionally be blocked by underwater obstacles. At the same time, an eavesdropper is assumed to monitor the acoustic hop, creating a secrecy concern. To address this, we formulate the relay power allocation problem as an infinite-horizon Markov decision process (MDP). A model-based reinforcement learning (RL) driven optimal power allocation (OPA) strategy is proposed to maximize  long-term cumulative secrecy performance until the network stops functioning. To offer lower-complexity alternatives, we also develop a Greedy Algorithm (GA) and a Naive Algorithm (NA). Simulation results show that the RL-based OPA adapts effectively to battery dynamics, varying channel conditions, and optical link availability, achieving the highest secure data transmission, while GA performs reasonably and NA performs poorly due to its short-sighted decisions.

\end{abstract}
\begin{IEEEkeywords}
Acoustic, energy harvesting, optical, reinforcement learning, secrecy capacity, underwater.
\end{IEEEkeywords}
\IEEEpeerreviewmaketitle
\vspace{-.2cm}


\section{Introduction}

Underwater wireless communication is becoming increasingly important for applications such as environmental monitoring, underwater surveillance, offshore exploration, and autonomous vehicle coordination. Among existing technologies, underwater acoustic (UWA) communication remains dominant due to its long transmission range, though it is limited by narrow bandwidth, high latency, and strong multipath effects \cite{intro_uwa, Octavia2019}. In contrast, underwater optical (UWO) communication enables much higher data rates with lower latency but is highly vulnerable to absorption, scattering, and physical blockage \cite{UOC_ICC25}. Since underwater nodes typically operate with limited or no external power supply, energy harvesting (EH) has become a practical approach to sustain long-term underwater network operation \cite{ku2015EH_paper}. Thus, in dynamic underwater environments, transmit power allocation becomes a key design aspect, as it affects link reliability and overall network lifetime. 

Future underwater networks are expected to operate more intelligently, with nodes adapting to environmental changes and system constraints to optimize long-term performance \cite{octavia2020VTC}. Reinforcement learning (RL) provides a suitable framework for such adaptive optimization, as it enables an agent to learn effective actions through interaction with the environment and reward-driven feedback \cite{sutton1998RL}. Unlike conventional optimization approaches that require accurate system models or future knowledge, RL can learn near-optimal strategies online under uncertainty and time-varying conditions \cite{sutton1998RL, puterman2014markov}.



Existing studies have explored RL-based power control in UWA relay networks. In \cite{Octavia2019}, a finite-horizon MDP is used to optimize the end-to-end sum rate in a three-node full-duplex EH relay system, solved via backward induction. Similarly, \cite{octavia2020VTC} employs an infinite-horizon MDP and value iteration to maximize long-term energy efficiency through adaptive relay power control. However, these works do not consider security threats, even though acoustic wireless signals remain vulnerable to interception by unauthorized users. In \cite{UWA_secrecy_Octavia_TVT}, secrecy performance in a full-duplex acoustic relay network using power-domain NOMA is analyzed in the presence of a mobile eavesdropper. Power allocation problems are formulated for scenarios with and without eavesdropper CSI and solved using convex optimization. In \cite{UWA_secrecyRL25}, secure UWA communication is enhanced using RL-assisted reconfigurable intelligent surface (RIS) optimization, where Q-learning and SARSA are used to adjust transmitter and RIS parameters in the presence of eavesdroppers. In \cite{UOC_ICC25}, an IRS-assisted underwater optical communication system is studied, where an ASV communicates with an AUV under frequent LoS blockage. The link selection problem is modeled using RL, and a deep Q-network is employed to determine the optimal switching policy.

While existing works address acoustic and optical systems separately, practical underwater environments require greater flexibility. Optical links provide high data rates but are vulnerable to blockage and turbulence, whereas acoustic links offer reliable range but lower capacity and higher interception risk. A hybrid optical–acoustic approach can better balance secrecy, reliability, and efficiency under dynamic underwater conditions. Motivated by these considerations, we study a hybrid underwater EH network with a source, an EH-enabled relay, a destination, and an eavesdropper. The source uses an optical link to send data to the relay, which forwards it acoustically using limited battery energy harvested according to a Bernoulli process. The system operates in discrete time slots, and its lifetime is random due to potential physical or hardware failure. Our objective is to allocate relay transmit power to maximize the long-term expected number of securely transmitted bits before the network stops operating. 
The key contributions are listed as follows:
\vspace{-.1cm}
\begin{itemize}
\item We formulate the relay power allocation problem in a hybrid underwater EH system with secrecy constraints, where the objective is to maximize the long-term expected number of securely transmitted bits before network termination. 
\item The problem is modeled as an infinite-horizon MDP, and the optimal power allocation (OPA) strategy is proposed using the policy iteration (PI) algorithm.
\item To provide lower-complexity alternatives, we propose two suboptimal schemes, (i) greedy algorithm (GA) and (ii) a naive algorithm (NA), and evaluate their secrecy performance.
\item We analyze the computational complexity of all schemes and demonstrate through numerical results that the proposed model-based RL driven OPA method achieves superior long-term secure throughput compared to GA and NA.
\end{itemize}



\section{System Model}

 \begin{figure}
 \raggedleft
  \includegraphics[width=3.4in]{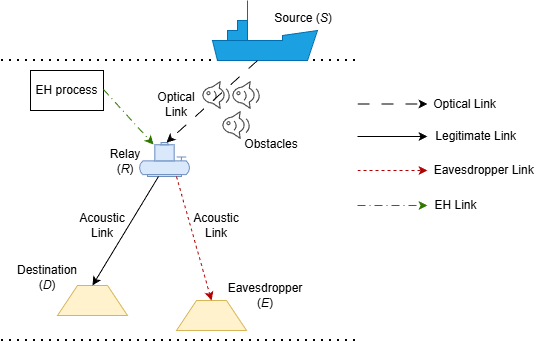}
 \caption{System model for the considered underwater system.}
 \label{fig_system}
 \vspace{-0.4cm}
\end{figure}

We consider a hybrid underwater optical-acoustic communication system comprising a surface vehicle source ($S$), a relay node ($R$), a destination node ($D$), and a passive eavesdropper ($E$), as shown in Fig. \ref{fig_system}. $S$ communicates with $R$ using UWO link and $R$ relays the information to $D$ using UWA link. $R$ is equipped with  EH device with limited energy storage, while $S$ relies on a traditional power source. Given the broadcast nature of UWA signal, $E$ overhears the communication from $R$ to $D$. The communication occurs in discrete time-slots (TSs), indexed by $k \in \mathcal{K} = \{0,1, \ldots, K-1\}$, where $K$ is the total number of available TSs and each TS has a fixed duration of $T_s$ seconds~\cite{dohler2013learning}. Since EH nodes may fail due to physical damage or hardware faults, the network lifetime $K$ is modeled as a Geometric random variable~\cite{dohler2013learning} with mean $1/(1-\Gamma)$, where $\Gamma\in[0,1)$ denotes the probability of network being functional in a given TS. Node $S$ is assumed to have an uninterrupted data supply for transmission in all TSs and node $R$ operates as a DF relay. All nodes works in half-duplex mode.


\subsection{EH Model} 
\vspace{-.1cm}
Harvested energy at $R$ in the $k$th TS is denoted as $H_R^{(k)} \in \mathcal{H}_R$. Here, $\mathcal{H}_R = \{0, E_R\}$ represents the set of feasible harvested energy levels. The EH process is modelled as a Bernoulli process~\cite{shalini_TVT} with probability $p$ and assumed to be independent of data transmission. Consequently, in each TS $k \in \mathcal{K}$, the probability of $R$ harvesting $E_R$ units of energy is given by $\mathbb{P}[H_R^{(k)}=E_R]=p$ while no energy is harvested with probability $\mathbb{P}[H_R^{(k)}=0]=1-p$. Node $R$ is equipped with finite-capacity battery having $B_{R}^{\max}$ energy units at max. The battery energy level at $R$ in the $k$-th TS is denoted by $B_R^{(k)} \in \mathcal{B}_R$, where $\mathcal{B}_R = \{0,1, \ldots, B_R^{\max}\}$ define the sets of possible discrete battery states.   

The energy used for data transmission in any TS is constrained by the available battery energy at $R$ in that TS. Additionally, the harvested energy cannot be stored beyond the battery's maximum capacity. The harvested energy $H_R^{(k)}$, together with the transmitted energy at the $k$-th TS, is utilized to update the battery state for the $(k+1)$-th TS as
\begin{align}
\label{battery_Bs}
   B_R^{(k+1)} 
   \hspace{-0.1cm}
   &= 
   \hspace{-0.1cm}
   \left\{ 
    \hspace{-0.2cm}
   \begin{array}{lcl}
    \min
    \{ B_R^{(k)} - P_R^{(k)}  T_s + E_R,B_R^{\textrm{max}} \}
    \,\, \mbox{for}\,\,H_R^{(k)} \hspace{-0.1cm}=\hspace{-0.1cm} E_R  
    \\ B_R^{(k)} - P_R^{(k)} T_s 
   \hspace{2.6cm} \mbox{for}\,\,H_R^{(k)} = 0 
    \end{array}
    \right\},
\end{align}
where $P_R^{(k)} \in \mathcal{P}$ represents the power levels used by $R$ for transmission in the $k$-th TS. The set of $M$ possible transmit power levels is given by $\mathcal{P} = \{P_1,P_2, \ldots, P_M\}$, ensuring that power constraint $P_R^{(k)} T_s \leq B_R^{(k)}$ holds. 
The EH in the $k$th TS will be available for use in the $(k+1)$-th TS onward. 
\vspace{-0.1in}
\subsection{UWO Channel Model for the $SR$ Link}
The UWO link is modelled as a composite fading channel and the received electrical signal at $R$ in $k$-th TS is
\begin{equation}
    y_R^{(k)} = \sqrt{P_S^{(k)}G_{SR}^{(k)}} \eta_o  s^{(k)} + n_o^{(k)},
\label{eq:rx_signal}
\end{equation}
where $P_R^{(k)}$ is the source power at $S$, $\eta_o$ is the photodetector responsivity, $s^{(k)} \in \{0,1\}$ denotes the transmitted optical intensity, and $n_o^{(k)}$ is AWGN sample at $R$ with variance $\sigma_N^2$. The composite channel gain is modeled as
\begin{equation}
    G_{SR}^{(k)}=  |h_a^{(k)} h_t^{(k)} h_p^{(k)} h_b^{(k)}|^2,
\label{eq:composite_gain}
\end{equation}
where $h_a^{(k)}$, $h_t^{(k)}$, $h_p^{(k)}$, and $h_b^{(k)}$ represent attenuation, turbulence, pointing error, and binary obstruction, respectively. The UWO attenuation $h_a^{(k)}$ follows Beer-Lambert's law as $h_a^{(k)} = \exp{(-c(\lambda)l_{SR})}$, where $l_{SR}$ is the $SR$ link distance and $c(\lambda)=a(\lambda)+b(\lambda)$ is the wavelength-dependent attenuation coefficient. Turbulence-induced fading $h_t^{(k)}$ follows a Gamma-Gamma distribution with PDF \cite{UOC_ICC25}
\begin{equation}
    f_{h_t^{(k)}}(x)= \frac{2(\alpha\beta)^{(\alpha+\beta)/2}}
    {\Gamma(\alpha)\Gamma(\beta)}
    x^{\frac{\alpha+\beta}{2}-1}
    K_{\alpha-\beta}\!\left(2\sqrt{\alpha\beta x}\right),
\label{eq:gamma_gamma}
\end{equation}
where $\alpha$ and $\beta$ are derived from the scintillation index. $\Gamma(\cdot)$ and $K_{\nu}(\cdot)$ denote the standard gamma function and second kind modified Bessel function of $\nu$-th order, respectively. The pointing error coefficient $h_p^{(k)}$ is modeled as
\begin{equation}
    f_{h_p^{(k)}}(x) = 
    \frac{\rho}{A_0^\rho} x^{\rho-1}, \qquad 0 \le x \le A_0,
\label{eq:pointing_error}
\end{equation}
where $\rho = w_{eq}^2/\sigma_p^2$, $w_{eq}$ is the equivalent beam radius, 
and $\sigma_p$ is the jitter variance. Obstruction due to underwater obstacles is captured via a binary coefficient $h_b^{(k)} \in \{0,1\}$ with $\Pr[h_b^{(k)} = 1] = e^{-T_o l_{SR}}$, where $T_o$ denotes the obstacle density. Taking into account the composite channel gain $G_{SR}^{(k)}$, the instantaneous SNR at $R$ is
\begin{equation}
    \gamma_{R}^{(k)} = 
    \frac{\eta_o^2 P_S^{(k)}}{\sigma_N^2} \, G_{SR}^{(k)}.
\label{eq:snr}
\end{equation}

\subsection{UWA Channel Model for $RD$ and $RE$ Links}
\vspace{-.1cm}
The UWA channel is modeled by frequency-dependent
attenuation and ambient noise. The power attenuation over distance \(l\) at frequency $f$ is
\begin{equation}
    A(l,f)=l^{k_s} a(f)^{\,l},
\end{equation}
where \(k_s\) is the spreading factor and \(a(f)\) is the Thorp absorption coefficient. Ambient noise is modeled as \cite{stojanovic2007relationship}
\begin{equation}
  N_{a}(f)=N_t(f)+N_s(f)+N_w(f)+N_{\rm th}(f),  
\end{equation}
with each component given in \(\mu{\rm Pa}/{\rm Hz}\). Let
\(N_{\rm dB}(f)=10\log_{10}N_{a}(f)\) is noise in dB re \(\mu{\rm Pa}/{\rm Hz}\). The corresponding electrical noise PSD (in W/Hz) is \cite{Octavia2019}
\begin{equation}
    N_W(f)=10^{\left(N_{\rm dB}(f)-120-10\log_{10}(\rho c)\right)/10} A_{\rm rec},
\end{equation}
where \(\rho\) is the water density, \(c\) is the sound speed in water, and \(A_{\rm rec}\) is the effective receiver aperture. Considering transmission over the band \([f_{\min},f_{\min}+B]\) in $k$-th TS, node $R$ transmits power $P_R^{(k)}$ with flat PSD \(S^{(k)}(f)=P_R^{(k)}/B\).
For acoustic $RX$ link, \(X\in\{D,E\}\), the broadband SNR at node $X$ in $k$-th TS is
\begin{equation}
\gamma_X^{(k)}=
\frac{{P_R^{(k)}}G_{RX}^{(k)}\displaystyle\int_{f_{\min}}^{f_{\min}+B} A^{-1}(l_{RX},f)\,df}
{{B}
\displaystyle\int_{f_{\min}}^{f_{\min}+B} N_W(f)\,df },
\end{equation}
where $G_{RX}^{(k)}$ is the small scale fading gain at $k$-th TS \cite{stojanovic2007relationship}. The values of $G_{RX}^{(k)}$ is quantized into $L$ discrete levels such that $G_{RX}^{(k)}  \in \mathcal{G}$, where $\mathcal{G} = \{G_1, G_2, \ldots, G_L\}$. Within a given TS, the channel power gain remains constant but transitions to a new value in the next TS according to the predefined set $\mathcal{G}$. This transition is assumed to follow a first-order Markov process \cite{dohler2013learning}, which models the uncertainty of UWA channels.

\subsection{Performance Metric}
\vspace{-.1cm}
We define the achievable secrecy rate in the $k$-th TS, measured in bits per second (bps), as the difference between the achievable rates of the legitimate and eavesdropping links as
\begin{align}
    C_S^{(k)} = \max \{C_D^{(k)} - C_E^{(k)},0\}\quad \text{bps},
    \label{eqCSk}
\end{align}
where $ C_D^{(k)} = B \log_2{(1+\min\{\gamma_D^{(k)}, \gamma_R^{(k)}
\})} $ and $ C_E^{(k)} = B \log_2{(1+\min\{\gamma_E^{(k)}, \gamma_R^{(k)}
\})}$ are the achievable rates for the end-to-end legitimate and eavesdropping links in the $k$-th TS, respectively. When the UWO link is unobstructed, it's SNR is much higher than that of the UWA link. Hence, $\min\{\gamma_X^{(k)}, \gamma_R^{(k)}
\}\approx\gamma_X^{(k)}$ indicating that the instantaneous secrecy rate $C_S^{(k)}$ is mainly limited by the UWA channel and by the obstacle coefficient in UWO channel, $h_b^{(k)}$. Therefore, $C_S^{(k)}$ can be approximated as
\begin{equation}
C_S^{(k)}\approx h_b^{(k)}C_{Sa}^{(k)},
\end{equation}
where $C_{Sa}^{(k)}=\max\left\{\log_2{\left(\frac{1+\gamma_{D}}{1+\gamma_{E}}\right)},0
\right\}$ is the secrecy rate of the UWA link. The expected total transmitted secure bits   until the network stops functioning is defined as \cite{wong2014joint}
\begin{align}
    \mu =  \mathbb{E} \left[  \mathbb{E}_K \left[ \sum_{k=0}^{K-1}   C_S^{(k)} \right] \right]\, \text{bps}, \quad \mbox{for} \,\, C_S^{(k)}\geq R_{th},  
    \label{eqECS}
\end{align} 
 where $\mathbb{E}_{K}[\cdot]$ denotes the expectation with respect to the random variable $K$ and $\mathbb{E}[\cdot]$ denotes the expectation taken over all other relevant random variables, i.e., $G_{RD}^{(k)}$, $G_{RE}^{(k)}$, $H_{R}^{(k)}$. The constant $R_{th}$ is the rate threshold for specific QoS requirement.  

\section{Problem Formulation}
\vspace{-.1cm}
The goal of the considered EH-based underwater communication system is to maximize the expected cumulative number of securely transmitted bits per second. This is achieved by optimally selecting $P_R^{(k)}$ during each TS until the network becomes non-operational. Based on this objective, the power allocation strategy at $R$ can be formulated as the following optimization problem:
\begin{subequations}
\begin{align}
    \text{P1}:  \underset{\{P^{(k)}_R\}_{k = 0}^K }
    {\text{maximize}} \quad
     &  \mathbb{E} \left[  \mathbb{E}_K \left[ \sum_{k=0}^{K-1}   C_{Sa}^{(k)} \right] \right]\label{optimization_a}\\
     \text{s.t.}\,
     & (\ref{battery_Bs})\label{optimization_b}\\
     & 0  \leq P_R^{(k)} \leq \frac{B_{R}^{(k)}}{ T_s} \label{optimization_c}\\
     & C_{Sa}^{(k)} \geq R_\textrm{th}^{(k)}. \label{optimization_d}
\end{align}
\label{Problem}
\vspace{-0.2in}\end{subequations}


The relay’s battery level evolves across time slots based on its transmit power consumption and harvested energy, as defined in (\ref{optimization_b}). The relay’s transmit power constraints and the secrecy rate requirement at $D$ are given in (\ref{optimization_c}) and (\ref{optimization_d}), respectively. Since the system satisfies the Markov property, the optimization problem in (\ref{Problem}) can be modeled as a sequential decision process with finite states, actions, and rewards. This naturally leads to an MDP formulation, which enables selecting the optimal action at each time slot to maximize the expected long-term reward \cite{puterman2014markov, bellman1957markovian}.


To apply RL, we first define the components of the MDP: decision epochs, state space, action space, reward, and transition probabilities \cite{sutton1998RL}. The decision epochs correspond to the TSs $k \in \mathcal{K}$. The system state at TS $k$ is represented as $s^{(k)} = (G_{RD}^{(k)}, G_{RE}^{(k)}, B_R^{(k)})$, and the overall state space is $\mathcal{S} = \mathcal{G}{RD} \times \mathcal{G}{RE} \times \mathcal{B}_R$, consisting of $N_S$ discrete states. At each TS $k$, the relay selects an action $a^{(k)}$ to solve problem P1, which corresponds to choosing a transmit power level $P_R^{(k)}$ from the feasible set $U(s^{(k)})$, such that $a^{(k)} \in U(s^{(k)})$, where
\begin{align}
\label{eq_feasible_action}
      U(s^{(k)}) = \bigg\{P_R^{(k)} \mid 0 \leq P_R^{(k)} \leq \frac{B_{R}^{(k)}}{T_s}\bigg\}. 
\end{align}
The action selected in the $k$-th TS, denoted by $a^{(k)}$, is taken from the set 
$\mathcal{A} = \{\delta_1, \ldots, \delta_{N_A}\}$ of all possible actions. Each action $\delta_i$, for $i \in \{1, \ldots, N_A\}$, corresponds to one of the relay's 
available transmit power levels $P_m \in \mathcal{P}$, where $m \in \{1, \ldots, M\}$. Since every power level represents a unique action, we have $N_A = M$. The probability of transitioning to the next state $s^{(k+1)}$ in the $(k+1)$-th TS from the current state $s^{(k)}$, after choosing action $a^{(k)}$, is expressed as
\begin{align}
    \mathbb{P}[s^{(k+1)}\mid s^{(k)}, a^{(k)}]
    &= \mathbb{P}[G_{RD}^{(k+1)}\mid G_{RD}^{(k)}] \times \mathbb{P}[G_{RE}^{(k+1)}\mid G_{RE}^{(k)}] \nn \\
    &\hspace{-0.6in}\times \mathbb{P}[B_R^{(k+1)} \mid B_R^{(k)}, H_R^{(k)}, P_R^{(k)}]
    \times\mathbb{P}[H_R^{(k)}].
    \label{Transition_Prob}
\end{align} 
Here, $\mathbb{P}[B_R^{(k+1)} \mid B_R^{(k)}, H_R^{(k)}, P_R^{(k)}]$ is equal to 1, whenever (\ref{battery_Bs}) is satisfied, and is zero otherwise. The harvesting process follows a Bernoulli model, where 
$\mathbb{P}[H_{R}^{(k)} = E_R] = p$ and $\mathbb{P}[H_{R}^{(k)} = 0] = 1 - p$. 
If $\mathbb{P}[B_R^{(k+1)} \mid B_R^{(k)}, H_R^{(k)}, P_R^{(k)}]$ evaluates to zero, then the transition probability in (\ref{Transition_Prob}) also becomes zero, indicating that moving from state $s^{(k)}$ to $s^{(k+1)}$ under action $a^{(k)}$ is not feasible. Each action $a^{(k)}$ also yields an immediate reward denoted by $R^{(k)}(s^{(k)},a^{(k)})$. The instantaneous reward in the $k$-th TS, as obtained from (\ref{eqCSk}), is given by
\begin{equation}
    R^{(k)}(s^{(k)},a^{(k)}) = 
    \left\{
    \begin{array}{cc}
        C_{Sa}^{(k)} & C_{Sa}^{(k)}\geq R_{th} \\
         0 & C_{Sa}^{(k)}<R_{th} 
    \end{array}
    \right\}.
    \label{reward}
\end{equation}
and the long-term expected total reward is expressed in (\ref{eqECS}). 
\vspace{-0.1cm}
\section{Proposed Solutions}
\vspace{-.1cm}
In this section, we present an optimal solution to the problem in (\ref{Problem}) using the policy iteration (PI) method \cite{puterman2014markov}, referred to as the OPA scheme. We also introduce two lower-complexity sub-optimal strategies as the Greedy and the Naive Algorithms. Details of all three methods are provided in the following subsections.
\vspace{-.1cm}

\subsection{Optimal Power Allocation (OPA)}
\vspace{-.1cm}

Algorithm~\ref{planning_phase} presents the pseudo-code of the OPA method. The process of generating the lookup table is referred to as the \textit{planning phase}, and the resulting policy is later used online to determine the relay's transmit power in each TS. The planning phase consists of two iterative steps: \textit{policy evaluation} and \textit{policy improvement}. In the policy evaluation step, the value function $V(s)$ is updated using the Bellman equation (line 6) until convergence (line 9). Then, in the policy improvement step (line 13), the policy $\hat{d}(s)$ is updated by selecting the action $a \in U(s)$ that maximizes the value function. These steps repeat until the policy stops changing, which provides the optimal solution.


\begin{algorithm}[t]
\caption{\textbf{Algorithm 1: The Planning Phase}}
\hspace*{\algorithmicindent} \textbf{Input}: Set of states, actions, state transition probability, and reward;\\
\hspace*{\algorithmicindent} \textbf{Output}: Optimal stationary deterministic policy $d^*(s)$ 
\begin{algorithmic}[1]
\State Initialize $V(s)$ and stationary deterministic policy $d(s)$ arbitrarily for all $s\in \mathcal{S}$, set small threshold $\epsilon$.

\begin{flushleft}
\textbf{Policy evaluation:} 
\end{flushleft}
\Repeat
\State $\Delta = 0 $ 
    \ForEach {$s \in \mathcal S $} 
    \State $v = V(s)$
    \State $V(s) =   [R(s,a) + \Gamma \underset{s^{'} \in \mathcal{S}} \sum \mathbb{P}(s^{'} \mid s, a) V(s^{'})]$ 
    \State $\Delta = \max(\Delta, |v- V(s)|)$
    \EndFor
\Until $\Delta < \epsilon $

\begin{flushleft}
 \textbf{Policy improvement:}
 \end{flushleft}
\State policy-stable $=$ true
    \ForEach {$s \in \mathcal S$} 
    \State $\hat{d}(s) = d(s)$
    \State $d(s) = \mathop{\mathrm{argmax}}\limits_{a \in U(s)} 
        \left[ R(s,a) + \Gamma 
        \sum_{s^{'} \in \mathcal{S}} \mathbb{P}(s^{'} \mid s, a) V(s^{'}) \right]$ 
        \If {$\hat{d}(s) \neq d(s)$}
        \State policy-stable $=$ false
        \EndIf
    \EndFor

\begin{flushleft}
\textbf{Check stopping criteria: } 
\end{flushleft}
\If {policy-stable}
\State stop
\Else
\State go-to policy evaluation (line-2) 
\EndIf
\end{algorithmic}
\label{planning_phase}
\end{algorithm}


The second stage of operation is the \textit{transmission phase}, where the relay uses the lookup table generated during the planning phase. The pseudo-code is shown in Algorithm~\ref{transmission_phase}. At each TS, the system determines the current state $s^{(k)}$ based on the channel and battery conditions, and then retrieves the corresponding optimal action $a^{(k)}$ from the table (line 7). This gives the transmit power $P_R^{(k)}$ for that TS. The cumulative reward and battery level are then updated accordingly.

\begin{algorithm}[t]
\caption{\textbf{Algorithm 2: Transmission Phase}}
\hspace*{\algorithmicindent} \textbf{Input}: Optimal stationary deterministic policy $d^*(s)$ and initial state $s^{(0)}$\\
\hspace*{\algorithmicindent} \textbf{Output}: Total expected discounted reward defined in (\ref{eqECS})
\begin{algorithmic}[1]
    \State Set $\mu = 0$
    \State Set $k = 0$
    \While{$k \leq K-1$}
    \State Track channel states $G_{RD}^{(k)}$ and $G_{RE}^{(k)}$ 
    \State Track available battery $B_R^{(k)}$ 
    \State Set $s^{(k)} = (G_{RD}^{(k)}, G_{RE}^{(k)}, B_R^{(k)})$
    \State Obtain $a^{(k)} = P_R^{(k)}$ from look-up table for state $s^{(k)}$
    \State Consume $P_R^{(k)}$ for transmission at relay.
    \State Calculate the total expected discounted reward $\Gamma^{k} C^{(k)}_S$ when $C^{(k)}_S\geq R_{th}$ for state $s^{(k)}$.
    \State Update battery $B_R^{(k)}$ using (\ref{battery_Bs})
    \State $\mu  = \mu  + \Gamma^{k}C^{(k)}_S $
    \State Set $k = k + 1$
    \EndWhile
\end{algorithmic}  
\label{transmission_phase}
\end{algorithm}

\subsection {Greedy Algorithm (GA)}
\vspace{-.1cm}
The GA algorithm does not involve a planning phase. Instead, during the transmission phase, 
it directly selects the action $a^{(k)} = \{P_{R}^{(k)}\}$ in each TS from the set of 
feasible actions $U(s^{(k)})$ for the current state $s^{(k)}$ that yields the highest 
instantaneous reward in (\ref{reward}) \cite{Octavia2019}. Based on this, the power allocation 
problem can be written as
\begin{align}
     a^{(k)} = \mathop{\mathrm{argmax}}\limits_{a^{(k)} \in U(s^{(k)})} R^{(k)}(s^{(k)}, a^{(k)}). 
     \label{GA_algorithm}
\end{align}

\subsection {Naive Algorithm (NA)}
\vspace{-.1cm}
The NA algorithm similarly does not require a planning phase. During the transmission phase, 
it simply uses all the energy available in the battery at node $\textrm{R}$ for transmission in each TS \cite{ahmed2012power, ahmed2013joint}. In this case, the transmit power at $R$ in the $k$th TS is given by $P^{(k)}_R = \frac{B^{(k)}_R}{T_s}$.

\subsection{Complexity Analysis}
\vspace{-.1cm}
The computational complexity of the proposed OPA scheme consists of two parts. The planning phase, where the optimal policy is computed, has a complexity of $\mathcal{O}\big(\frac{N_A^{N_S}}{N_S}\big)$ \cite{PIcomplexity}, while the transmission phase, which simply retrieves actions from the precomputed lookup table, has a complexity of $\mathcal{O}(K)$ \cite{shalini_TVT}. In contrast, the GA does not require any planning phase. Its computational effort lies solely in the transmission stage, where it evaluates all possible actions at each time slot, resulting in a complexity of $\mathcal{O}(K N_A)$ \cite{Ajib_VTC}. Similarly, the NA operates without a planning stage and requires minimal computation during transmission, yielding a complexity of $\mathcal{O}(K)$ \cite{ahmed2013joint}.

\section{Results and Discussions}
\vspace{-.1cm}
In this section, we compare the secrecy performance of the OPA, GA, and NA schemes in terms of the expected total number of securely transmitted bits. Unless specified otherwise, the simulation parameters are configured as follows. Each time slot has a duration of $T = 1$ second. For the UWO link, the parameters are considered as: $\lambda = 520$ nm, $c(\lambda)=0.15$/m, Gamma--Gamma fading parameters $\alpha = 5.54$ and $\beta = 3.92$, equivalent beam waist $w_{eq} = 0.1$ m, pointing error jitter $\sigma_p = 0.01$, aperture coefficient $A_0 = 0.05$ m, source transmit power $P_S = 10$~W, obstacle density $T_0 = 10^{-4}$, and source-to-relay distance $l_{SR}=120$ m. For the UWA link, we set $f_{\min} = 9.5$ kHz and bandwidth $B = 5$ kHz, with relay--destination and relay--eavesdropper distances of $l_{RD} = 5$ km and $l_{RE} = 6$ km, respectively. The parameters $k$, $w_s$, and $s$ are set to $2$, $0$, and $0.5$. The relay selects its transmit power from $P_R \in \{0, 1, 2, 3\}$ W. The channel gains are drawn from the discrete set $G_{RX} \in \{0.106,\, 0.511,\, 1.61\}$, with the transition matrix:
\[
{\small
\begin{bmatrix}
0.82 & 0.18 & 0 \\
0.09 & 0.81 & 0.10 \\
0 & 0.09 & 0.91
\end{bmatrix}}
\]

The relay battery is discretized into five levels with a maximum capacity of $B_{R}^{\max} = 5$ W. The discount factor is set to $\epsilon = 0.001$, and energy harvesting is modeled with harvested energy $E_R = 2$ W and harvesting probability $p = 0.6$. The initial system state is selected as $s^{(1)} = (G_3,\, G_3,\, B_{R}^{\max})$.

\begin{figure}[t!]
  \centering  
  \includegraphics[width=0.8\columnwidth]{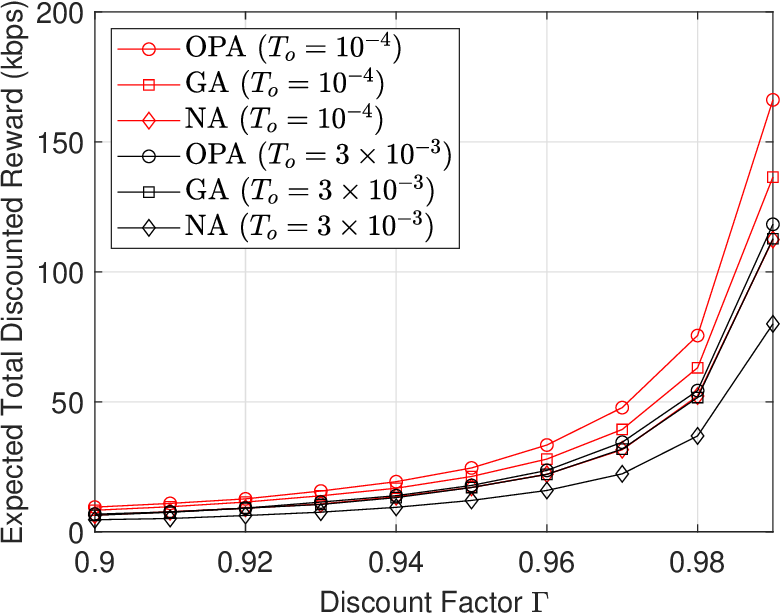}
  \caption{Expected total discounted reward versus discount factor $\Gamma$ with different obstacle density $T_o$.}
  \label{fig_ETD_vs_Gamma}
\vspace{-0.2in}\end{figure}

In Fig.~\ref{fig_ETD_vs_Gamma}, we evaluate the expected total discounted reward as a function of $\Gamma$ for the OPA, GA, and NA schemes under different obstacle densities, where $T_o$ varies from $3\times10^{-3}$ to $10^{-4}$. The results indicate that increasing $\Gamma$ leads to a higher expected total discounted reward for all three schemes. This trend is expected since a larger $\Gamma$ places greater emphasis on future rewards, encouraging power allocation strategies that are more efficient over the network's operational lifetime. Among the evaluated schemes, the OPA algorithm consistently achieves the highest reward. This performance gain stems from its ability to incorporate both current and future system conditions when selecting transmit power. The GA demonstrates moderate performance since it optimizes only the instantaneous reward at each TS without considering the long-term impact. In contrast, the NA provides the lowest reward because it neither adapts to system dynamics nor accounts for immediate or future reward variations. Another important observation from Fig.~\ref{fig_ETD_vs_Gamma} is that increasing obstacle density leads to a noticeable degradation in performance across all algorithms. This reduction occurs because a higher obstacle density reduces the reliability of the optical link between the source and relay, resulting in fewer successful secure transmissions and, consequently, lower accumulated reward.

\begin{figure}[t!]
  \centering  
  \includegraphics[width=0.8\columnwidth]{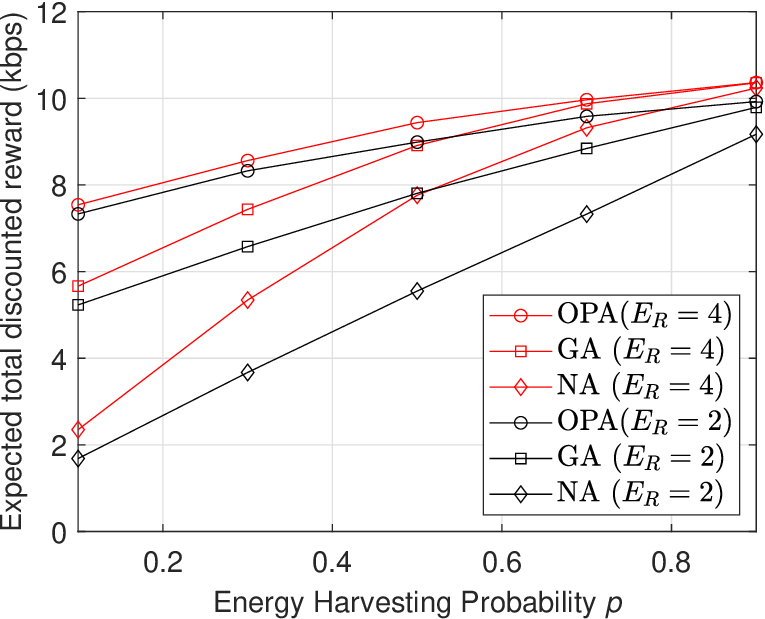}
  \caption{Expected total discounted reward versus EH probability $p$ when $E_R$ improves from $2$ to $4$.}
  \label{fig_ETD_vs_EH_varyp}
\vspace{-0.2in}\end{figure}

Fig.~\ref{fig_ETD_vs_EH_varyp} illustrates the expected total discounted reward as a function of the EH probability $p$, comparing the OPA, GA, and NA algorithms for harvested energy levels $E_R \in \{2,3,4\}$. The results show that the reward increases for all algorithms as $p$ grows, since a higher harvesting probability ensures more frequent energy availability at the relay, enabling more secure transmissions before the network ceases operation. Another key observation is that the performance gap between OPA, GA, and NA decreases at higher values of $p$. When energy becomes consistently available in the battery, the benefit of long-term optimal planning diminishes, causing all algorithms to perform similarly. Additionally, increasing the harvested energy amount $E_R$ improves performance for all cases, as more energy per harvesting event further supports secure transmission. Overall, the results confirm that both higher EH probability and larger harvested energy values lead to better secrecy performance by providing a more reliable energy supply at the relay.

\begin{figure}[t!]
  \centering  
  \includegraphics[width=0.8\columnwidth]{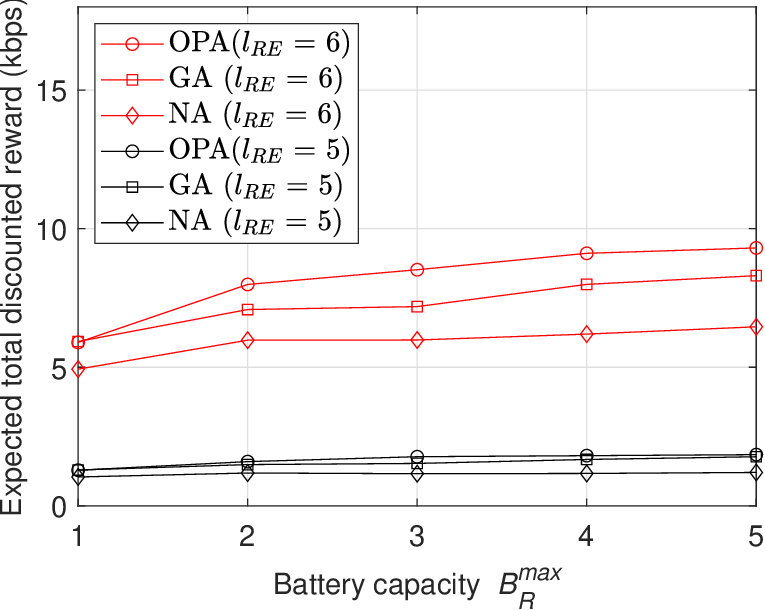}
  \caption{Expected total discounted reward versus battery capacity $B_{R}^{\textrm{max}}$ when $l_{RE}$ changes from $5$ km to $6$ km.}
  \label{fig_ETD_vs_Brmax}
\vspace{-0.2in}\end{figure}

Figure~\ref{fig_ETD_vs_Brmax} presents the expected total discounted reward versus the relay battery capacity $B_{R}^{\max}$ for different relay–eavesdropper distances $l_{RE}$, comparing the OPA, GA, and NA schemes. The results show that increasing $B_{R}^{\max}$ improves performance for all algorithms, as a larger battery allows the relay to store more harvested energy and use it later for secure transmission, thereby extending operation and increasing the number of securely transmitted bits. Additionally, reducing the $RE$ link distance leads to a noticeable performance degradation. A shorter $l_{RE}$ strengthens the eavesdropping channel, reduces secrecy capacity, and limits secure throughput. Overall, the figure demonstrates that both sufficient battery storage and a weaker eavesdropping channel are crucial for maximizing secrecy performance in energy-harvesting underwater networks.

\section{Conclusion}
In this paper, we analyzed a secure underwater communication system where a surface source communicates with an underwater destination via an energy-harvesting relay in the presence of an eavesdropper. The relay power allocation problem was modeled as an infinite-horizon MDP, and an OPA strategy based on policy iteration was developed. Two simpler benchmark schemes, GA and NA, were also evaluated. Simulation results show that the proposed OPA approach achieves the best secrecy performance by considering long-term energy and channel dynamics, while GA performs moderately and NA yields the lowest performance.


\section*{Acknowledgement}
This work was supported in part by Taighde Éireann – Research Ireland under Grant number 22/PATH-S/10788.
 
\bibliographystyle{IEEEtran}
\bibliography{IEEEabrv,PA_RL}
\vspace{1cm}

\end{document}